\documentclass{osa-article}

\journal{osajournal}

\articletype{Research Article}
\usepackage{listings}
\usepackage{tikz}
\usepackage{tabularx}
\usepackage{booktabs}

\newcommand{\higherit}[1]{\overset{\xleftarrow{\text{Higher Iterations}}}{#1}}

\newcommand{\MyTikzmark}[2]{\tikz[overlay,remember picture,baseline] \node [anchor=base] (#1) {$#2$};}

\newcommand{\DrawVLine}[3][]{ \begin{tikzpicture}[overlay,remember picture] \draw[->, #1] (#2.north) -- (#3.south); \end{tikzpicture}}

\begin{document}
\title{On-Chip Optical Convolutional Neural Networks}

\author{Hengameh Bagherian,\authormark{1,*} Scott Skirlo,\authormark{1} Yichen Shen,\authormark{1,$\dagger$} Huaiyu Meng, \authormark{2} Vladimir Ceperic, \authormark{1,3} and Marin Solja\v{c}i\' c\authormark{1}}
\address{\authormark{1}Massachusetts Institute of Technology, Department of Physics, Cambridge, MA 02139, USA\\
\authormark{2}Research Laboratory of Electronics, Massachusetts Institute of Technology, Cambridge, MA 02139, USA\\
\authormark{3}Faculty of Electrical Engineering and Computing, University of Zagreb, Unska 3, 10000 Zagreb, Croatia\\
}
\email{\authormark{*}hengameh@mit.edu}
\email{\authormark{$\dagger$}ycshen@mit.edu} 



\begin{abstract}
Convolutional Neural Networks (CNNs) are a class of Artificial Neural Networks (ANNs) that employ the method of convolving input images with filter-kernels for object recognition and classification purposes. In this paper we propose a photonics circuit architecture which could consume a fraction of energy per inference compared with state of the art electronics.
\end{abstract}

\section{Introduction}
Exploration of neuromorphic computing architectures began in the late 1950s with the invention of the perceptron, which functioned as a binary classifier with a linear decision boundary \cite{rosenblatt1958perceptron}. The preceptron worked well for certain tasks, but further progress was hindered by a lack of understanding on how to handle multilayer versions. Progress on neuromorphic computing for image processing accelerated rapidly in the 1990s, when LeCun et al. pioneered using back-propagation on an architecture based on convolving images with kernels, known as Convolutional Neural Networks (CNNs) \cite{lecun1998gradient,hinton2006reducing,goodfellow2016deep}. This architecture consists of successive layers of convolution, nonlinearity, downsampling, followed by fully connected layers (see Fig. 1a). The key to the success of CNNs was that convolution and downsampling handled the translation invariance of image features efficiently, while the multiple layers allowed greater flexibility in training than the few-layer approaches.
\par
Although the CNN architecture successfully managed to implement digit classification at human performance levels and compared favorably to other machine learning techniques, it was not until improvements in processing speeds and the creation of large human-labeled image databases from the Internet, that the full potential of CNNs became apparent \cite{ILSVRC15}. Using GPU-accelerated backpropagtion, AlexNet achieved record breaking results on ImageNet for a thousand categories using a CNN architecture composed of five convolutional layers and three fully connected layers \cite{krizhevsky2012imagenet,ILSVRC15}. Following AlexNet's lead, modern CNNs of dozens or hundreds of layers, and hundreds of millions to billions of parameters, can achieve better than human level performance in many image classification tasks \cite{he2015delving,he2016deep}. Recent breakthroughs with DeepLearning such as playing Atari games \cite{mnih2015human}, by combining reinforcement-learning and CNNs, have convinced many that these networks are some of the best tools for a new machine learning golden age with applications ranging from pedestrian detection for self-driving cars to biomedical image analysis \cite{lecun2015deep,schmidhuber2015deep,silver2016mastering,benenson2014ten,esteva2017dermatologist,yu2016predicting,
li2015convolutional}.
\par
A big part of this success story was the advent of GPU-acceleration for large matrix-matrix multiplications, which are the essential and most time intensive step of back-propagation in CNN training. Despite significant gains, training large CNNs takes weeks utilizing large clusters of GPUs. More practically, GPU-accelerated CNN inference is still a computationally intensive task, making image analysis of the vast majority of the image and video data generated by the Internet very difficult. Youtube itself, in 2015 experienced uploads of 300 hours of video every minute \cite{brouwer2014youtube}; this would require a cluster of 18000 Nvidia Titan X GPUs to process continuously with CNNs, drawing 4.5 Megawatts, with the hardware costing tens of millions of US dollars \cite{kang2017optimizing}. 


\par
Given that this is just one company and that video traffic is predicted to grow to be 80$\%$ of the Internet by 2020 \cite{cisco2012cisco}, this problem is going to get harder and will far outpace the current computing paradigms, requiring investment in specialized neuromorphic \emph{hardware} architectures. There are many proposals and experimental demonstrations to accomplish this through analog circuits, digital ASIC designs, FPGAs, and other electronic technologies \cite{mead1990neuromorphic,poon2011neuromorphic,horowitz20141,shafiee2016isaac,misra2010artificial,merolla2011digital,chen2016eyeriss}.

\par
Our work follows a long history of optical computing such as optical implementations of unitary matrix multiplication, optical memory, all optical switching, optical interconnects, and even recent works on optical neuromorphic architectures such as photonic spike processing and reservoir computing \cite{nozaki2010sub,rios2015integrated,sun2013large,sun2015single,tanabe2005fast,
shen2016deep,tait2014photonic,tait2014broadcast,prucnal2016recent,vandoorne2014experimental}. We focus primarily on integrated photonics as a computation platform because it provides the highest raw bandwidth currently available of any technology that is mass manufacturable and has standardized components.

\begin{figure}[!]
\centering\includegraphics[width=\textwidth]{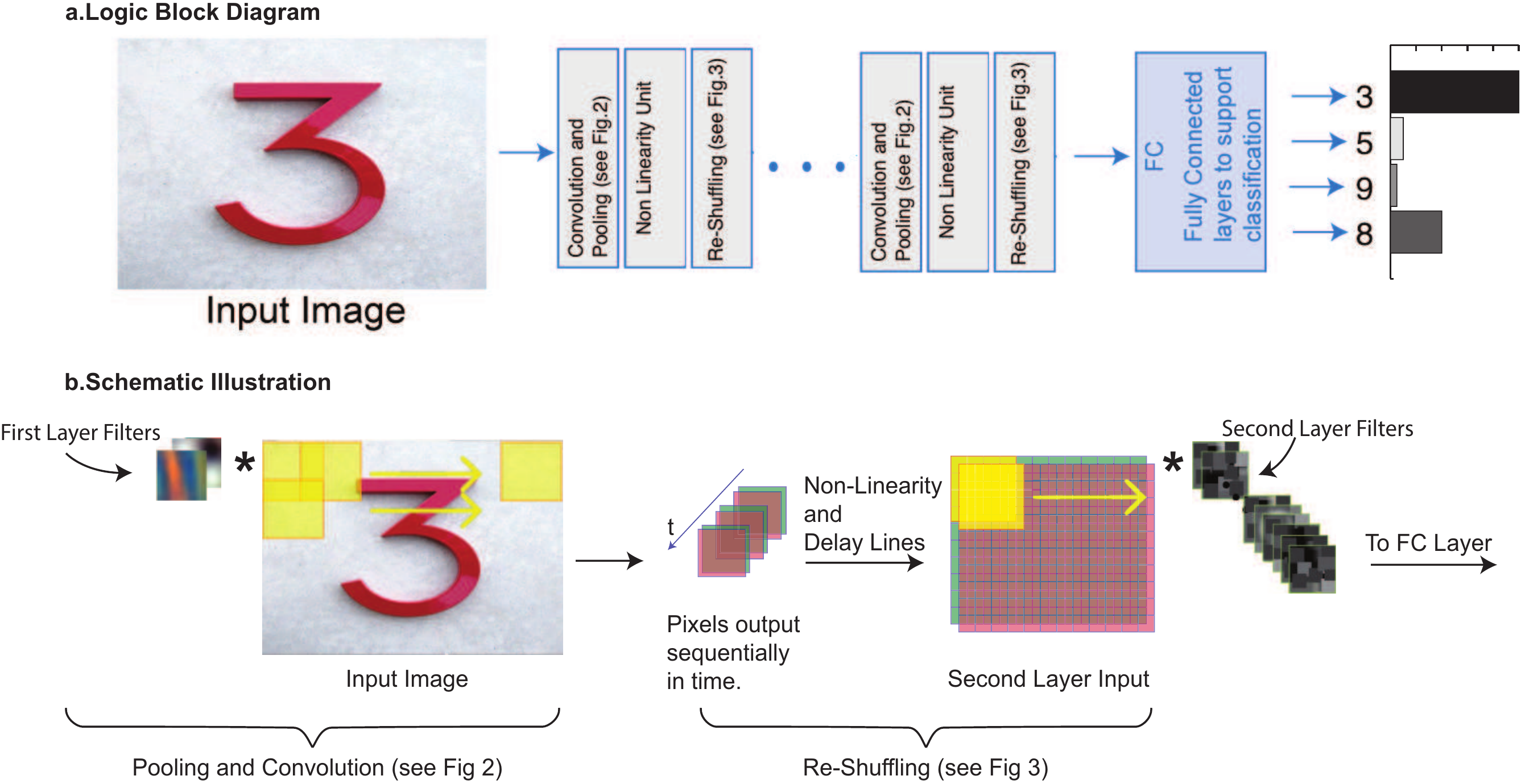}
\caption{Convolutional Neural Net (CNN) Architecture. a. Logic Block Diagram: The input image, number 3 shown here, is passed through successive layers of convolution and pooling, nonlinearities (see Fig. 2 for further description ), and re-shuffling of the pixels (see Fig. 3 for further description). A final fully connected layer maps the last stage of convolution output to a set of classification outputs. b Schematic Illustration: First part of CNN implements convolution of the image with a set of smaller filters. These produce a sequence of kernel-patch dot products which are passed through a nonlinearity and are re-shuffled into a new d-dimensional image, where d is the number of filters in the first layer. The process is then repeated on this new image for many subsequent layers.}
\end{figure}

\section{Architecture}

\par
As depicted in the block diagram form in Fig. 1a, and pictorially in Fig. 1b, the CNN algorithm consists of several main steps, each of which we need to execute optically. First the image is convolved with a set of kernels. The output is a new image with dimensions $[(W-K)/S+1]\times[(W-K)/S+1]\times[d]$, where $W$ is the original image width, $K$ is the kernel dimension, $S$ is the convolution stride and $d$ is the number of kernels. Next, the new image is subject to pooling, where the image produced by convolution is further downsampled by selecting the maximum value of a set of pixels (max pooling) or taking their average (average pooling). After pooling, a nonlinearity is applied to each pixel of the downsampled image. This nonlinearity can consist of the rectified linear unit (ReLU), sigmoid, tanh, or other functions. Following nonlinearity, the entire process is repeated with new sets of kernels and the same nonlinearity. After a number of these convolution layers, a fully connected neural network is applied to the output to perform the final processing steps for classification.


\par
We begin our discussion of an optical kernel convolution, with the description of a related GPU algorithm \cite{chetlur2014cudnn}. One of the chief advantages of utilizing GPUs for machine learning is their ability to execute large matrix-matrix multiplications rapidly. For fully connected neural networks, it is obvious how this capability can lead to large speedups. To see how this works for CNNs, we first depict the conversion of an image into a set of ``patches'', the same dimension as the kernels in Fig. 2a. In a GPU algorithm, these patches can be converted into a ``patch matrix'' by vectorizing and stacking each patch. The patch matrix can then be efficiently multiplied by a ``kernel matrix'', formed by vectorizing and stacking each kernel. The output is a matrix composed of kernel-patch dot products which can then be ``re-patched'' for multiplication by the next layer's kernel matrix.

\par
A recent work \cite{shen2016deep} utilized networks of MZIs and variable loss waveguides for optical matrix multiplication for fully connected neural networks. Here we propose using such a network of MZIs and variable loss waveguides to implement the kernel matrix multiplication, as depicted in Fig. 2b. The patch matrix takes the form of a sequence of coherent optical pulses whose amplitude encodes the intensity of each patch pixel from an image. In turn, each output of the photonic circuit will correspond to a time series of Kernel-patch dot products with the amplitude and phase ($0$ or $\pi$) of each pulse encoding the output of the computation. Fig. 2b depicts this process schematically, where the photonic circuit outputs at different times have been drawn in their corresponding locations in the next layer's image. Furthermore, because MZIs 
take up a large chip space, by doing this time multiplexing method, we are minimizing the number of MZIs we use to be equal to the number of degrees of freedoms in the kernal matrices -- the theoretical limit for any hardware implementation using weight-stationary data flow scheme \cite{chen2016eyeriss} \footnote{Another natural way of carrying out convolutions in optical domain is using lens-spacial-light-modulator(SLM)-lens system. The advantage of such a system is that one can feed in the entire image in one clock (instead of feeding through multiple patches in our system). However, the disadvantage for such approach is SLM modulation speed is 6 orders of magnitude slower than an integrated modulator, and for each kernel one would need a separate lens-SLM-lens system. For a typical CNN, each layer contains hundreds of kernels, which will effectively make the system to be bulky.}

\par
In addition to implementing convolution with kernel matrices, the photonic circuit in Fig. 2 is also providing two other functions: pooling and nonlinearity. Pooling (or downsampling) is realized by taking the stride of the convolution to be greater than one. This step occurs when the patches are formed from the input image. This provides a necessary information filtering function required to reduce the image to only a few bits of information identifying its class. CNNs working this way have been shown to have the same performance as those using max or average pooling \cite{springenberg2014striving}. 

Finally, optical or electrical nonlinearity and repatching is applied to each output of the circuit, providing the remaining ingredient for a complete CNN layer. In the following section, we will discuss two different ways of carrying out such nonlinearity and repatching, and compare their performances.


\begin{figure}[!h]
\centering\includegraphics[width=\textwidth]{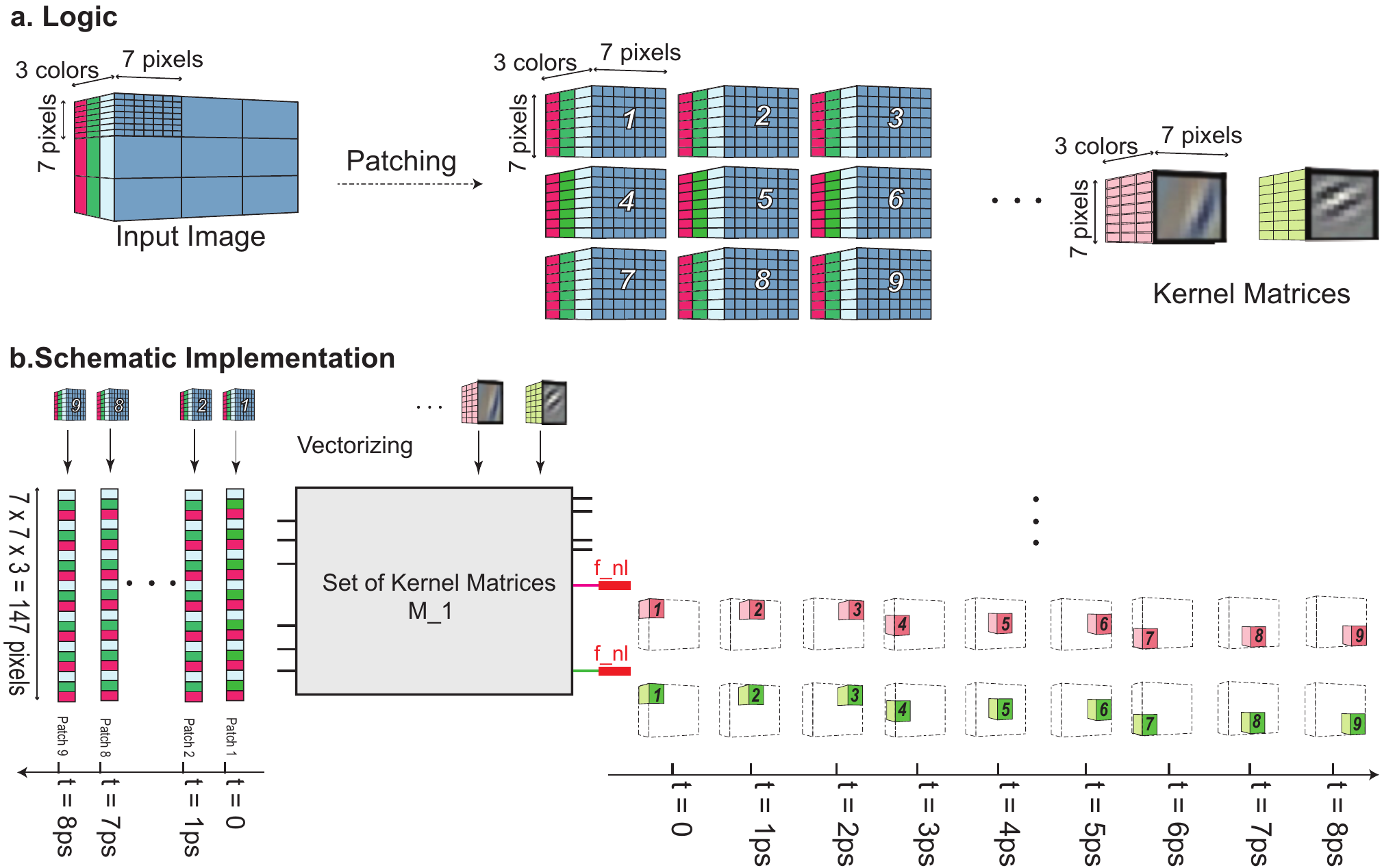}
\caption{General Optical Matrix Multiplication a. Logic: The pixels of the input image on the left (21 x 21 x 3 colors) are grouped into smaller patches, which have the same dimension as the kernels of the first layer ( depicted on the right-hand side). b. Schematic Implementation: Each of these patches is reshaped into a single column of data that is sequentially fed, patch by patch, into the optical interference unit. Signal propagation of the optical data column through the unit implements a dot product of the first layer kernels with the patch input vector. The result is a time series of optical signals whose amplitude is proportional to the dot products of the patches with the kernels. Each output port of the optical interference unit corresponds to a separate time series of dot products associated with a given kernel. Optical nonlinearity is applied to each output port of the optical interference unit. }
\end{figure}

\begin{figure}[!h]
\centering\includegraphics[width=\textwidth]{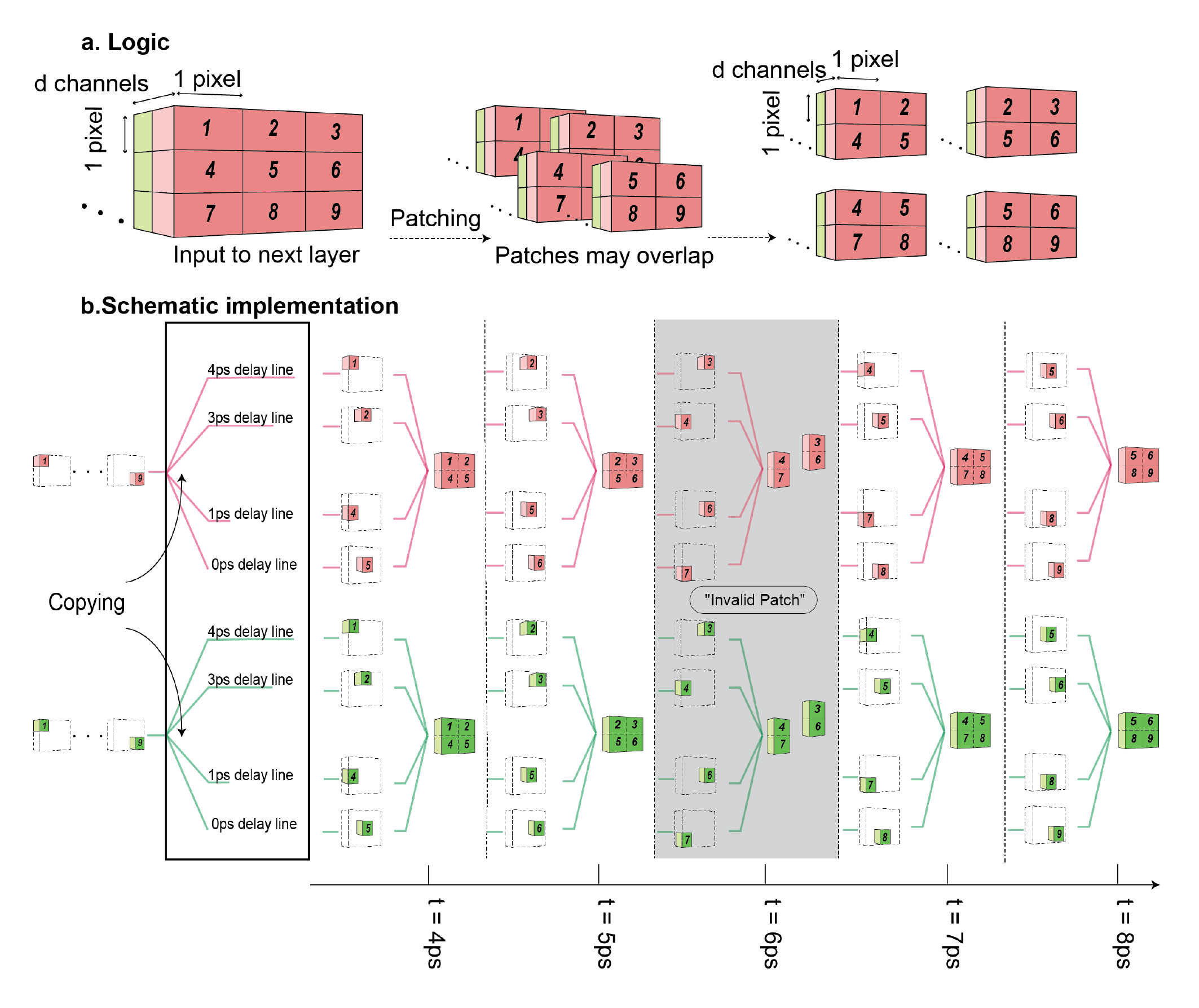}
\caption{Optical Delay Lines for repatching a. Logic: The output kernel dot products from the first layer (right hand side Fig2.b) are depicted as cubes on the left in Fig3. Each cube is labeled with the timestep the corresponding kernel dot product was computed. The right hand side indicates the repatching procedure necessary to convert the set of kernel dot products on the left hand-side into input patches the same size as the next layer's kernels. b. Schematic Implementation: The optical delay lines are designed such that a sequence of kernel dot products can be reshuffled in time to form a new patch the same size as the next layer's kernels. Each delay line is connected through 3-dB splitters to the original signal line, allowing the data to be copied and then delayed for synchronization.  The reshuffling procedure produces valid patches only at specific sampling times. The grayed out section at t=6ps indicates an invalid sampling interval, where the patches have partially wrapped around.}
\end{figure}

\section{Physical Implementation}



\par
\subsection{Full Optical Setup}
Figure 4 shows a full implementation of the envisioned architecture using optical matrix multiplication, optical nonlinearity and optical delay lines. To illustrate the fundamental features of the photonic integrated circuit more clearly we have omitted the required optical amplifiers for each convolution layer, and some parts of the optical matrix multiplication.

\par
The first part of the circuit consists of an optical interference unit. The matrix $M_{i}$ encodes the kernels for a given convolution layer. From SVD decomposition we know that $M_{i}=U\Sigma V$, where $V$ and $U$ are unitary and $\Sigma$ is some real diagonal matrix. $U$ and $V$ are implemented through Reck-encoding of the MZI matrix, and $\Sigma$ through tunable waveguide loss \cite{shen2016deep,harris2015bosonic,PhysRevLett.73.58,Metcalf16}. An incorrect realization of the unitary matrices (among many other possible errors) will degrade a network's inference capability, which we explore in the Appendix through a numerical example. In Fig. 4 we have omitted the optical implementations of $\Sigma$ and $V$ for simplicity.

\par
The next stage consists of optical nonlinearity applied to each output waveguide of the MZI matrix. As suggested in ref. 31, optical nonlinearity can be realized by using graphene, dye, or semiconductor saturable absorbers \cite{cheng2014plane,soljavcic2002optimal,schirmer1997nonlinear,bao2011monolayer,selden1967pulse}. Additionally ref. 31 showed that the nonlinear response of graphene saturable absorbers has a suitable functional form for training neural networks. The power budget required for operating AlexNet, utilizing a graphene saturable absorber nonlinearity operating at $\sim 0.05$mW per waveguide input powers is detailed in the Appendix B \cite{Bao2010,Li2012}.
\par
We find in the appendix there that a single inference can be computed with $P_{inf}$ power, which is given by:
\begin{equation}
1.26\cdot 10^{11} \Delta t P_{0}=P_{inf}
\end{equation}
Taking $P_{0}\sim 0.05$mW, $\Delta t=\frac{1}{f}=1/3$ ns, where $f=3$ GHz is the throughput (from the maximum delay-line bandwidth), we find that we require $2$ mJ per inference which is of the same order of magnitude as that of an electronic setup demonstrated in Appendix C. Furthermore the optical implementation is 30 times faster than GPU-enabled inference with AlexNet (See Appendix C for calculations).

\par
Finally a single convolution layer ends with repatching logic consisting of a tree of 3dB splitters feeding into variable length delay lines. 
We propose to re-patch through a set of optical delay lines and splitters. The requirements for the splitting and delay procedure can be understood from Fig. 3a. Here we depict the time sequence of kernel-patch dot products from Fig. 2b, as a single image, with each pixel of the image labeled with the timeslot associated with the computation. This image needs to be converted into four patches on the right of Fig. 3a, which will need to appear as a time sequence input for the next optical matrix multiplication. How this is accomplished is illustrated in Fig. 3b. Here a given output from the previous layer is split into four separate waveguides and subject to different delays. Each delay line is selected such that at a given time, the outputs from the previous layer are synchronized in time and form a new patch for input into the next layer's kernel matrix. Since in this particular example we are forming two by two patches, a delay line of one time unit is required for the top right signal to arrive at the same time as the top left signal. Further delay lines of three and four time units are required for the bottom left and right signals to arrive with these. We illustrate the formation of new patches on the right hand side of Fig. 3b, where at specific times we have formed the four desired patches from Fig. 3a. Note that the grayed out section indicates a sampling time when an invalid patch is formed, that is the wrong set of pixels have arrived simultaneously. Since the original length of the patch matrix input is nine time units long, and since there are only four patches for input into the next system, there will be five such invalid sampling times in the period of the original input signal. The exact length of the delay lines for each layer are determined (see Appendix E), and the maximum delay line length is $\sim 5000 \Delta t$. In this approach, the total power consumption is solely determined by the optical power needed to trigger optical nonlinearity, and ranges from a minimum value of 2mJ/image (a detailed power calculation is described in Appendix B).


\par
To get a general sense of the engineering difficulties, if one use this setup to carry out conventional convolutional neural networks, delay lines have been engineered $1$ ns long with a $3$ GHz bandwidth using $200$ ring resonators on $0.2$mm$^2$ area \cite{lenz2001optical,tanabe2007trapping,xia2007ultracompact,rasras2005integrated}. Assuming we are using this technology, and $\Delta t\sim 1$ns, this will require $5000$ such delay units or $5 \mu$s of delay. Since there are $256$ outputs for the final AlexNet layer \cite{krizhevsky2012imagenet}, we have to assume we would require at least this many maximum delay line chips with a total area of $\sim 500$cm$^{2}$. Engineering and integrating delay lines of such length is a substantial engineering challenge.





\begin{figure}[!h]
\centering\includegraphics[width=\textwidth]{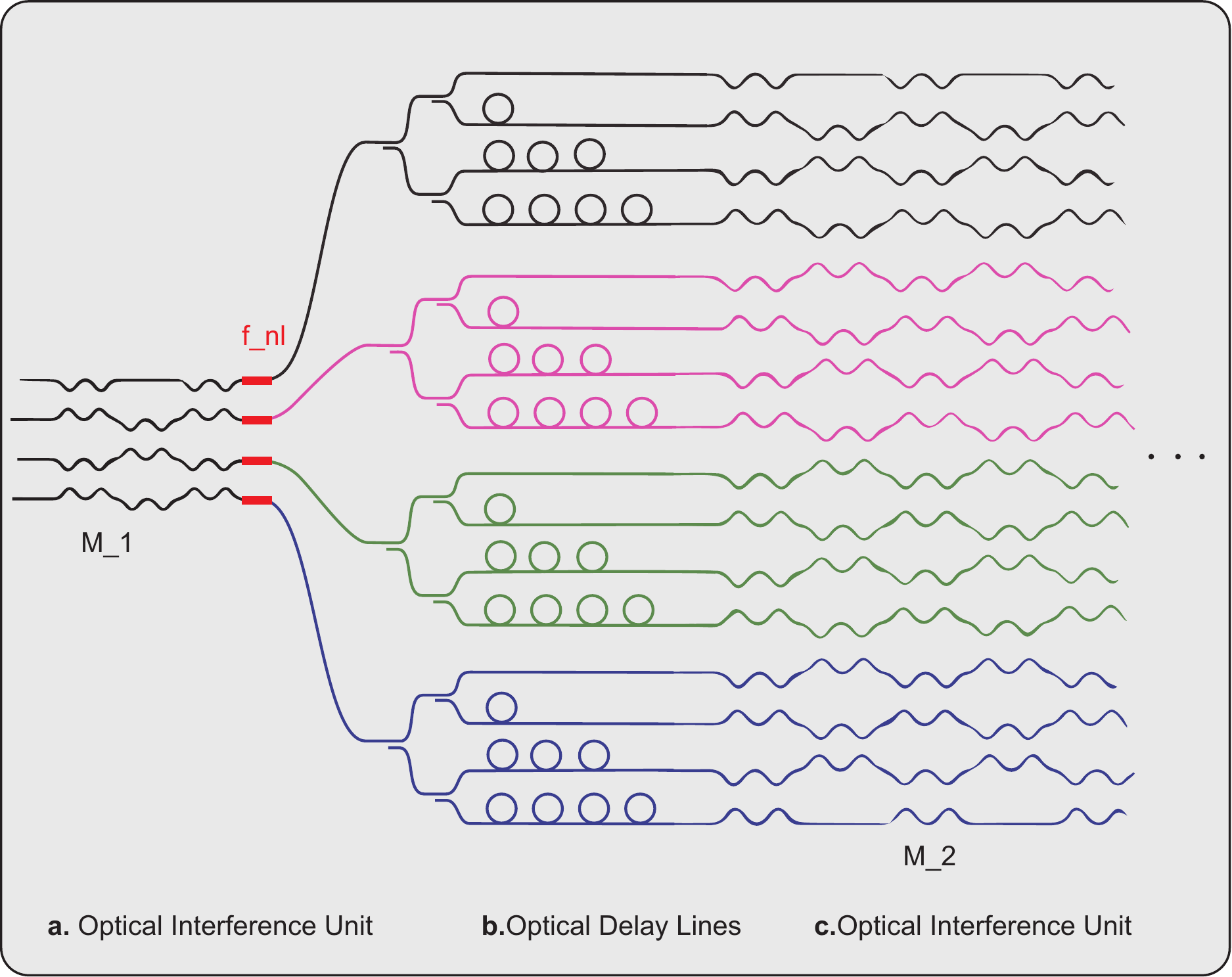}
\caption{Illustration of optical interference unit with delay lines a. Optical interference unit: In the first stage an optical interference unit is used to implement a kernel matrix $M_{1}$ which processes the patches from the original image. The red segments on the output of  $M_{1}$ are optical nonlinearity. b. Optical Delay Lines:  In the second stage, optical delay lines properly reform the sequence of kernel dot products into new patches for input into the second kernel matrix $M_{2}$. c.Optical interference unit: In the third stage the next optical interference unit is used to implement $M_{2}$ (partially depicted here). For clarity the actual number of inputs and outputs have been reduced and the attenuator stage and subsequent additional optical interference units have been omitted from $M_{1}$ and $M_{2}$. Additionally we have omitted optical amplifiers required in each layer which boost the power sufficiently to trigger the optical nonlinearity. }
\end{figure}

\par
\subsection{Optical-Electronic Hybrid Setup}
It is possible to replace the photonics circuits that are difficult to implement by their electronic counterparts. Due to the slow conversion of analog to digital signals and vice-versa, we propose that electronic counterparts operate in the analog domain (the schematic data flow of such system is illustrated in Fig. 5). Otherwise, the speed gained by photonics circuits could be offset by slow conversion. Furthermore, such conversion would require additional electronic circuits and overall power consumption could increase. The non-linearity needed in the convolution neural networks can be implemented by simple CMOS circuits such as \cite{Carrasco-Robles09,VrtaricCB13}. Since the voltage-current characteristic of a diode resembles ReLU, even a simple diode can be used as a nonlinear activation function. The long optical delay lines can be replaced by e.g. memristors \cite{Payvand14}, capacitance based analog memories \cite{Hock2013AnAD, Soelberg94ananalog}. The benefits of such hybrid architecture are reduced overall size, simpler implementation and on-chip integration of the system. A detailed analysis of future work on this system and its components should focus on increasing the bandwidth. This task is critical for maximizing the inference rate of the system, shortening the delay line length, and decreasing the power consumption per inference. However every component, including the optical nonlinearity, MZIs, optical detectors, and signal generators, can potentially limit the bandwidth. Currently we are limited in bandwidth by the present state-of-the-art compact delay lines \cite{xia2007ultracompact}. Beyond this, a significant limiting factor will be optical detectors which operate up to $120$ GHz \cite{vivien2012zero}. A detailed derivation of the energy consumed by the hybrid setup is found in Appendix D. 

\begin{figure}[!h]
\centering
\includegraphics[width=\textwidth]{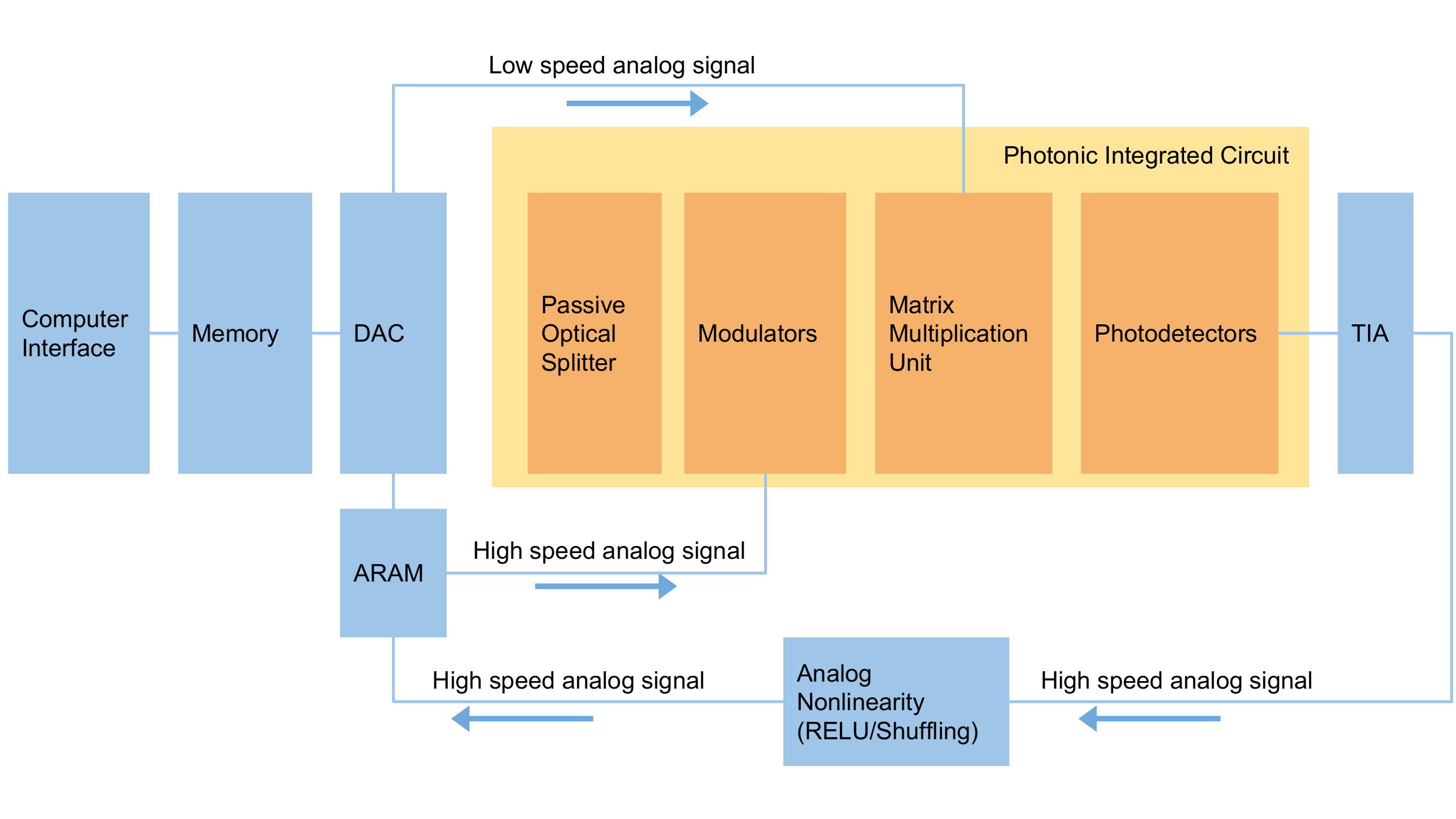}
\caption{The photonic integrated circuit (PIC) consists of four stages. An input laser is first split equally. Then the light will go through an array of optical amplitude modulators. Afterwards, the modulated optical signal will pass through the optical matrix multiplication unit (OMMU). Finally, the outputs are fed into an array of photodetectors and converted to photocurrents. On the electronic side, the system first receive data from an interface to external computational device and store them in a memory. The data are then converted to analog. The matrix weights are applied to the OMMU through a low speed link. Afterwards, the images information are transfered to a high speed analog memory (ARAM) before sent to optical modulators. The outputs of PIC are typically weak, so a transimpedance amplifier (TIA) array is used. A high speed ASIC circuit is used to perform shuffling and nonlinear operations, such as RELU or sigmoid. Afterwards, the rearranged images are sent back to the ARAM and ready for the next cycle of matrix multiplication. The image data will go through a large number of cycles in analog domain before being converted back to digital. Therefore, no high speed analog to digital conversion is necessary.}
\end{figure}

\section{Conclusion}
We have outlined the operation of a novel photonics integrated circuit that is capable of implementing high speed and low power inference for CNNs \footnote{As we were completing this work for publication, another article was posted on the Arxiv, suggesting a related optical CNN implementation \cite{mehrabian2018pcnna}.}. The system we outlined is related to a recently published work which investigated fully connected optical neural networks\cite{shen2016deep}. Here we bridge the gap between the fully connected optical neural network implementation and an optical CNN through the addition of precisely designed optical delay lines connecting together successive layers of optical matrix multiplication and nonlinearity. This new platform, should be able to perform on the order of a million inferences per second, at $2 $mJ power levels per inference, with the nearest state of the art ASIC competitor operating $30$ times slower and requiring the same total power \cite{chen2016eyeriss}. This system could play a significant role in processing the thousands of terabytes of image and video data generated daily by the Internet. Although the implementation of the photonics circuit proposed in this manuscript would be a substantial engineering challenge, the benefits of successfully realizing it are difficult to understate. In the meantime one could replace the photonic sub-circuits that are difficult to implement by their electronic counterparts.

\section{Acknowledgements}
We are grateful to Yann LeCun for his comments on an earlier version of this work, and encouraging us to look into this direction. We acknowledge useful discussions with Dirk Englund and John Joannopoulos. Research was sponsored by the Army Research Office and was accomplished under Cooperative Agreement Number W911NF-18-2-0048. 

\section*{Appendix A: Analysis of MZI phase encoding and error}

Although the analog nature of our optical CNN can allow for high precision computation, it also suffers significantly more than equivalent digital architectures from error propagation. There are many sources of error within our system that include variable waveguide loss, variable optical nonlinearity, variable amplification, and shot noise. We do not discuss these here, but for the purposes of providing a basic model of error propagation through the circuit, we calculate the classification error from incorrect phase settings of a Reck-encoded MZI matrix \cite{PhysRevLett.73.58,Metcalf16,miller2015perfect}.

To estimate how errors in MZI phase settings effect classification accuracy, a simulation was done on a toy model of a digit recognition CNN. Our toy CNN was comprised of two convolution layers and two fully connected layers and was trained on the MNIST digit recognition data set. After training the reference CNN (which had a classification accuracy of 97$\%$ on the test data set), the kernel matrix for each layer was exported to a phase extractor program that calculated the phase settings necessary to implement the unitary component of these matrices with MZIs (i.e. $U$ and $V$ from SVD decompostion of $M=U\Sigma V$).

We describe how we can calculate a phase encoding for any real valued unitary matrix. By applying a rotation matrix $T_{i,j}(\theta)$ to real unitary matrix $U$, with appropriate $\theta$ we can null elements in the $i$th row and $j$th column, where $T_{ij}(\theta)$, with $i=j-1$, is an identity matrix with elements $T_{ii},T_{ij},T_{ji},T_{jj}$ replaced by a two by two rotation matrix:

\begin{equation}
U' =
\left[
\begin{array}{cccc}
      1       &                                                               &                                                               &  0      \\ \cline{2-3}
               & \multicolumn{1}{|c}{\cos(\theta_{j-1,j})}  & \multicolumn{1}{c|}{\sin(\theta_{j-1,j})}  &          \\
               & \multicolumn{1}{|c}{-\sin(\theta_{j-1,j})}  & \multicolumn{1}{c|}{\cos(\theta_{j-1,j})} &          \\ \cline{2-3}
      0       &                                                               &                                                               &  1       \\
\end{array}\right]
\left[
\begin{array}{ccccc}
      u_{1,1}& \cdots  & u_{1,j-1}  &  0                                             & 0            \\  \cline{4-4}
                 & \cdot    &                & \multicolumn{1}{|c|}{u_{i,j}}      & \cdot      \\
                 &             & \cdot       & \multicolumn{1}{|c|}{u_{i+1,j}}  & \cdot      \\ \cline{4-4}
                 &             &                &  d_{k}                                       & 0            \\
                 &             &                &                                                 & d_N          \\
\end{array}\right]
\end{equation}

To determine $\theta$ in $T_{i,j}(\theta)$ such that the $ij$th element of the matrix is nulled, $\theta$ must satisfy the following equation:

\begin{equation}
\begin{split}
u'_{i,j} = \cos(\theta_{j-1,j})u_{i,j} + \sin(\theta_{j-1,j})u_{i+1,j} = 0\\
\tan(\theta_{j-1,j}) = \frac{-u_{i,j}}{u_{i+1,j}} \\
\end{split}
\end{equation}

If we apply these rotations starting with the first element of the far-right column and working downwards we find that:

\begin{equation}
T_{N,N-1} T_{N,N-2} \cdots T_{N,2} T_{N,1} U(N) =
\left[
\begin{array}{c|c}
     U(N-1)       & 0 \\
	               &\vdots \\ \hline
	0  \cdots &  a
    \end{array}
\right]
\end{equation}

Now we note that the block matrix $ U(N-1) $ can undergo the same process, and thus after $(N-1)+(N-2) \cdots= \frac{N(N-1)}{2} $ rotations the right hand side will turn into a diagonal real matrix $D$ (this is only true if $U$ is real, which it is for conventional CNNs). In total $U$ can be written in the following way \cite{PhysRevLett.73.58} \cite{Metcalf16}:
\begin{equation}
 U = (T_{2,1} T_{3,2} T_{3,1}T_{4,3} \cdots T_{N,N-1}\cdots T_{N,1})^{-1}D
\end{equation}

We can extract a matrix of phase encodings $\Theta$ with the following pseudocode:

\begin{lstlisting}[mathescape=true,language=Python]
function Phase_Extractor

for i from N to 2 :
	for j from 1 to i :
		$\Theta_{j,j+1} = \tan^{-1}( \frac{-U_{j,i}}{U_{j+1,i}} )$
		Update $ U = T_{j,j+1}(\Theta_{j,j+1})  U$	
	end for	
end for	
return $\Theta $
\end{lstlisting}

The algorithm nulls the elements of a given unitary matrix starting from $ U_{1,N}$ and moving downward until reaching a diagonal element, upon which it moves to the next column to the left. The below schematic shows the order in which the algorithm nulls the elements and extracts the corresponding phases:

\begin{equation*}
\higherit{\left[ \begin{array}{ccccc}
 d_1  & \MyTikzmark{TopA}{u_{1,2}} \MyTikzmark{bottomA}{u_{1,2}} & \cdots & \MyTikzmark{TopB}{u_{1,N-1}}          & \MyTikzmark{TopC}{u_{1,N}} \\
 	& d_2                                                                                           & \cdots & u_{2,N-1}                                            & u_{2,N} \\
 	&	                                                                                            & \ddots &\vdots                                                  &\vdots \\
	&	                                                                                            &            & \MyTikzmark{bottomB}{u_{N-2,N-1}} &\vdots \\
	&                                                                                                  &            & d_{N-1}                                              &\MyTikzmark{bottomC}{u_{N-1,N}} \\
	&                                                                                                  &            &                                                           &  d_N \\
      \end{array}\right]} \\
\end{equation*}

\DrawVLine[black,thick, opacity=0.5]{TopA}{bottomA}
\DrawVLine[black,thick, opacity=0.5]{TopB}{bottomB}
\DrawVLine[black,thick, opacity=0.5]{TopC}{bottomC}


Each element of the resulting phase matrix $\Theta$ is randomly perturbed with a distribution given by $p(\Delta \Theta_{i,j})=\frac{1}{\sqrt{2\pi \sigma}}\exp{\big[-\frac{\Delta \Theta_{i,j}^2}{2\sigma^2}\big]}$. This perturbed matrix is used to create a new $U^{\prime}$ (in $M=U^{\prime}\Sigma V^{\prime}$), which is in turn used to generate the kernel matrices of a perturbed CNN. The results of this perturbed network's digit classification on MNIST are then compared to those of the trained unperturbed CNN such that the network is assumed to be ``error free'' if it gives the same results as the reference network, not if it has 100$\%$ correct classification. These results are plotted in Fig. 6.

We find that for an error distribution with $\sigma > 0.01$, the performance of the perturbed network is significantly degraded relative to the unperturbed version. This corresponds to about 8-bit accuracy in the phase settings, which has been achieved in the previous work \cite{shen2016deep} on fully connected optical neural networks. These results are promising, but larger CNNs need to be examined with this method to assess their tolerance to the phase setting errors, and the other errors mentioned briefly above.

\begin{figure}[!h]
\centering\includegraphics[width=\textwidth]{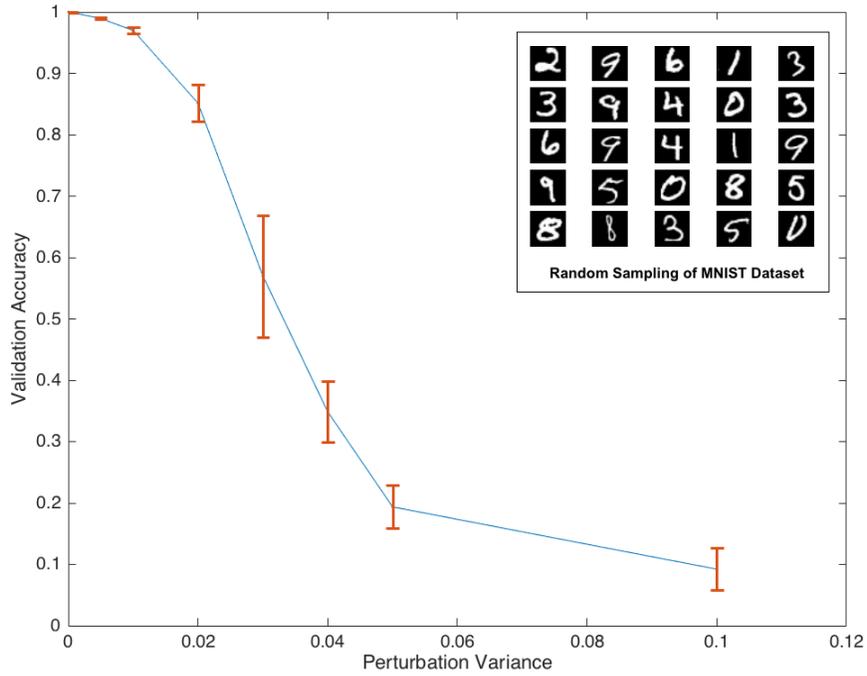}
\caption{Performance of digit recognition network on MNIST dataset vs. training parameter perturbation. After the matrix of phases, $\Theta$, has been returned by the phase extractor algorithm, the entries are perturbed and the resulting phase matrix is used in composing a new unitary matrix of weights that builds up a second CNN with desired perturbations. The perturbed CNN is tested and its inference performance on the MNIST dataset is then compared to those of the unperturbed CNN  to analyze the error.}
\end{figure}

\section*{Appendix B: Power consumption for optical implementation of AlexNet}

To ensure proper operation of the saturable absorption nonlinearity, the input power needs to be of the order of $P_0\sim 0.05$ mW per waveguide for graphene to trigger the nonlinear effect \cite{Bao2010,Li2012}. This requires that the output in each layer be amplified to that signal level, since the average power level is divided every layer by waveguide splitters in the repatching logic. To accommodate this, optical amplifiers are needed at every layer to boost the patch inputs appropriately. Given that standard integrated InP amplifiers operating in the IR have a wall plug efficiency of 10$\%$ or more \cite{juodawlkis2007advances}, this means that the power dissipated by the network will scale as $0.5 \frac{mW}{waveguide}$.

Since we are feeding $55\times 55=3025$ patch units at a rate of $\sim3$GHz  each optical nonlinearity unit requires $E=3025\cdot10\frac{P_{0}}{f} \approx 5 \cdot 10^{-10}\text{J/waveguide}$ for each signal. Consequently if we sum the number of waveguides in each layer and multiply by this number, we should get an idea of the power requirements for this type of integrated photonic circuit: \\

{\centering \small
\begin{tabularx}{\textwidth}{XXXXX}
\toprule
 & \# of ReLU units & \# of Input Patches & Layer Energy Consumption(J)\\
\midrule
 1st Conv &  96 & $55\times55$ & $1452000\times10^{-10}$\\
 2nd Conv & 256 & $55\times55$ & $3872000\times10^{-10}$\\
 3rd Cnv & 384 & $55\times55$ & $5808000\times10^{-10}$\\
 4th Conv & 384 & $55\times55$ & $5808000\times10^{-10}$\\
 5th Conv & 256 & $55\times55$ & $3872000\times10^{-10}$\\
 1st FC & 4096 & $1\footnote{The number of input patches decreases to one for the fully connected layers because we employ electro-optic converters to extract one valid patch from the convolutional layer output. In principle this technique could be employed to extract and re-emit only valid patches from previous layers, but we do not discuss this here, because we are primarily interested in an all-optical implementation.}$ & $20480\times10^{-10}$\\
 2nd FC & 4096 & 1 & $20480\times10^{-10}$\\
 3rd FC & 1000 & 1 & $5000\times10^{-10}$\\
 Total Energy Consumption &&& $20,857,960\times10^{-10}$\\
\bottomrule
\end{tabularx}} \\

, this yields a total energy consumption per inference of $2$ mJ for AlexNet \cite{krizhevsky2012imagenet}. We can further rewrite this as $1.26\cdot 10^{11}\Delta t P_{0}$.

\section*{Appendix C: Power consumption for electronic implementation of AlexNet}
Electronic computers consume a fixed amount of energy per floating point operation. Since data movement (i.e. data transfer between hard drive and RAM, etc.) represents additional significant ``overhead'' for memory intensive algorithms like CNNs, and minimizing this is the chief objective of new neuromorphic digital architectures, calculating the total number of floating point operations required for AlexNet gives us a good estimation of the best case performance for a digital implementation of that algorithm: \\

{\centering \small
\begin{tabularx}{\textwidth}{XXXXX}

 \toprule
  & Kernel Size &  Number of Input Patches & Number of Kernels & Layer FLOPs\\
  \midrule
 1st Conv & $11\times11\times3$ & $55\times55$ & 96 & $105415200\times2$\\
 2nd Conv & $5\times5\times96$ & $27\times27$ & 256 & $ 447897600\times2$\\
 3rd Conv & $3\times3\times256$ & $13\times13$ & 384 & $ 149520384\times2$\\
 4th Conv &$3\times3\times384$ & $13\times13$ & 384 & $224280576\times2$\\
 5th Conv & $3\times3\times384$ & $13\times13$ & 256 & $ 149520384\times2$\\
 1st FC & $13\times13\times256$ & $1\times1$ & 4096 & $177209344\times2$\\
 2nd FC & $ 1\times1\times4096$ & $1\times1$ & 4096 & $16777216\times2$\\
 3rd FC & $1\times1\times4096$ & $1\times1$ & 1000 & $4096000\times2$ \\
 Total FLOPs&&&& $2,549,433,408$\\
 \bottomrule
\end{tabularx}} \\

, electronic computers have an average performance rate of 
$112 \frac{\text{TFLOPs}}{\text{s}}$ at $250 \text{ W}$ \cite{NVIDIA}, so the lower bound on the total energy consumed by a digital computer (ASIC, GPU, CPU) running AlexNet is $ 2,549,433,408\text{ FLOPs} \times(112 \frac{\text{TFLOPs}}{\text{s}})^{-1} \times 250 \text{ W} \approx 5.7\text{mJ}$ \cite{krizhevsky2012imagenet}. In reality, the power consumption per image is a lot higher, because not all layers (e.g. fully connected layer) can be carried out at TPU's peak performance.

To compare the temporal performance of an electronic setup with that of a full optical implementation the time interval over which each circuit processes an image is calculated. For the electronic circuit, dividing the total number of FLOPs by the performance rate provides us with the amount of time it takes to process one image: $2,549,433,408\text{ FLOPs} \times(112 \frac{\text{TFLOPs}}{\text{s}})^{-1} \approx 2.2 \times 10^{-5} \text{s}$. Meanwhile, the optical counter part of the image processing time is derived from dividing the total number of patches within an image by the throughput of the system, $ 3025 \text{ unit patches} \times (3\text{GHz})^{-1} \approx 10^6 \text{s}$, revealing that the optical setup is $30$ times faster than its electronic counterpart.

\section*{Appendix D: Power consumption for optical-electronic hybrid implementation of AlexNet}

The bulk of the energy consumption in the optical-electric hybrid implementation is from analog electronics. Number of analog operations for each layer can be obtained by kernel size $\times$ number of input pads $\times$ 2.

{\centering \small
\begin{tabularx}{\textwidth}{XXXXX}

 \toprule
  & Kernel Size &  Number of Input Patches & Analog Operations\\
  \midrule
 1st Conv & $11\times11\times3$ & $55\times55$ & $1.10\times 10^6\times2$\\
 2nd Conv & $5\times5\times96$ & $27\times27$ &  $1.75\times 10^6\times2$\\
 3rd Conv & $3\times3\times256$ & $13\times13$ & $3.89\times 10^5\times2$\\
 4th Conv &$3\times3\times384$ & $13\times13$ & $5.84\times 10^5\times2$\\
 5th Conv & $3\times3\times384$ & $13\times13$ & $5.84\times 10^5\times2$\\
 1st FC & $13\times13\times256$ & $1\times1$ & $4.32\times 10^4\times2$\\
 2nd FC & $ 1\times1\times4096$ & $1\times1$ & $4096\times2$\\
 3rd FC & $1\times1\times4096$ & $1\times1$ & $4096\times2$ \\
 Total &&& $8.96\times 10^6$\\
 \bottomrule
\end{tabularx}} \\

Major source of power consumption in analog circuits are .0transimpedance amplifiers, optical modulator drivers and analog memories. Typical off-the-shelf components have 20 pJ/sample, 100 pJ/sample, respectively. Analog memory power consumption is estimated using typical CMOS capacitor:

\begin{align}
\text{Energy per bit}&=CV^2\times\text{number of capacitors}\nonumber\\
&=100fF\times (1V)^2\times 10^3\nonumber\\
&=100pJ
\end{align}

Therefore, the total analog energy for each image is around $(20+100+100)\times8.96\times 10^6=2$mJ/image .

Laser power can be calculated by:

\begin{align}
\text{Energy per bit}&=\text{laser power}\times\text{total number of patches} /\text{modulation speed}\nonumber\\
&=2W\times 4264 /10^9\nonumber\\
&=8.5\mu J
\end{align}

This is negligible compared with analog electronic energy consumption.

Recent research \cite{wade_energy-efficient_2014} shows that, with monolithic integration of electronic circuits into the photonic network, it is possible to dramatically reduce the energy consumption. Modulators cost as little as 5fJ/sample. Additionally, the analog memory could potentially be designed in such a way that only a small fraction of the data are moved for each cycle. As a result, the total energy consumption in the optical-electronic hybrid implementation can be reduced substantially. The energy consumption for each image could be as low as 45nJ/image.

\section*{Appendix E: Delay Line Length Calculation}
The calculation to derive each delay line length begins with the following definitions:
\begin{description}
  \item[Column Delay, $\Delta t_j$ $\coloneqq$] The delay time required for two adjacent pixels in the same row to arrive at the same time.
  \item[Row Delay, $\Delta T_j$ $\coloneqq$] The delay time required for two adjacent pixels in the same column to arrive at the same time.
\end{description}
An example of Column delay is the delay line, with $\Delta t_j = 1ps$, required for the top right signal to arrive at the same time as the top left one on the right of Fig. 3a.
The delay line, with $\Delta T_j = 3ps$, required for the top right signal to arrive at the same time as the bottom right one on the right of Fig. 3a. is an example of a row delay.
\par Looking at Fig. 7 one can write the following recursive formulas for row and column delay lengths:

\begin{equation}
\begin{split}
\Delta t_j &= \Delta t_{j-1} \times S_{j-1}\\
\Delta T_j &= \Delta T_{j-1} \times S_{j-1}
\end{split}
\end{equation}

And the maximum delay line length for each layer can be calculated as :

\begin{equation}
\text{Max Delay Length} = (K_j-1)\times(\Delta T_j + \Delta t_j)
\end{equation}

\begin{figure}[!]\centering\includegraphics[width=\textwidth]{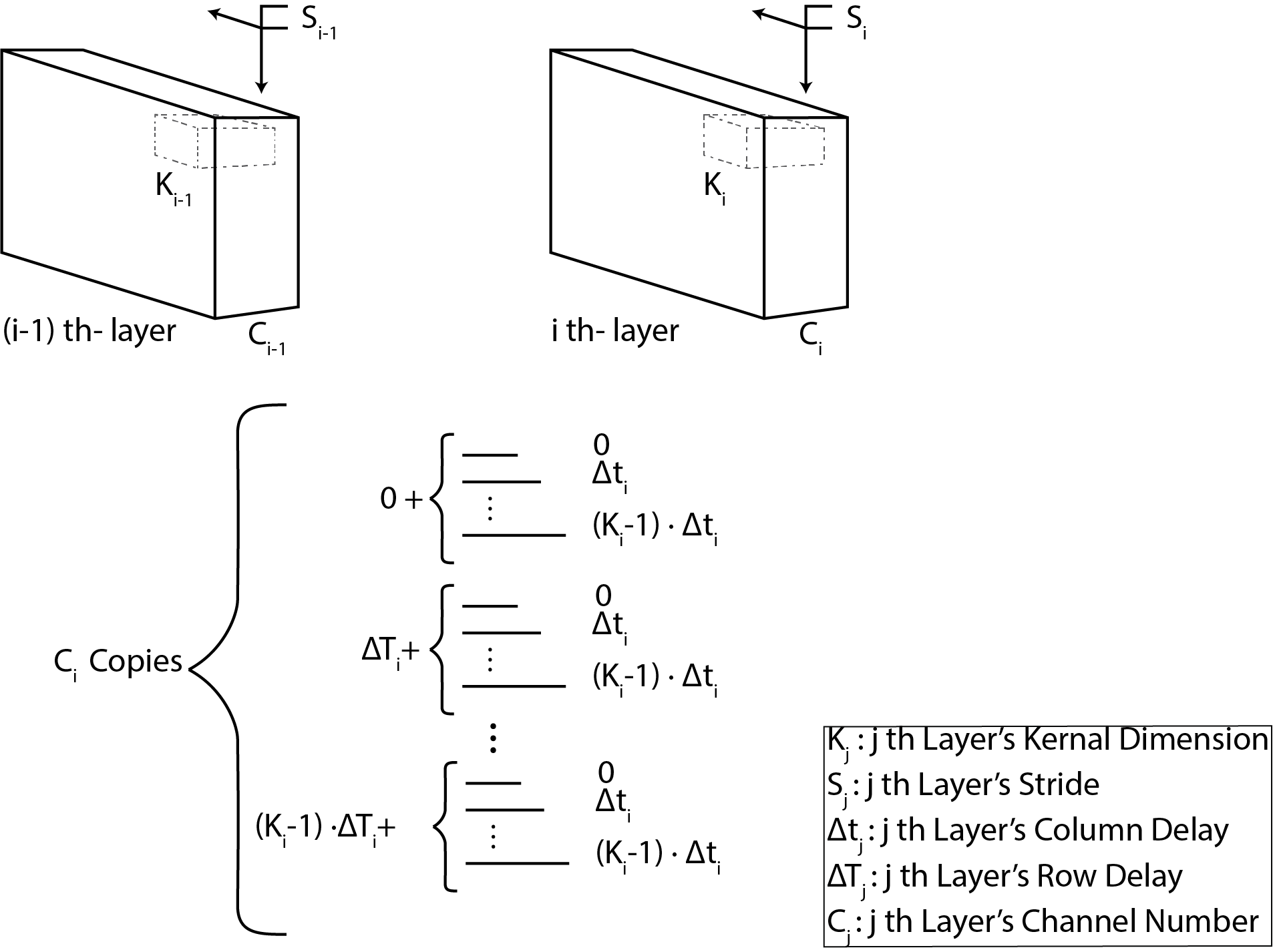}
\label{fig:figure_1}
\caption{Delay lines connecting the $(i-1)th$ and $i \ th$ layers of a CNN.}
\end{figure}

The initial values for Column Delays, $\Delta t_0$, and Row Delays, $\Delta T_0$, are derived using the properties of the input image and the first convolution layer. Assuming that the input image has dimensions of $ W \times W$, zero-padding of $P$, and is being fed into the system with a frequency $f$ we get :

\begin{equation}
\begin{split}
\Delta t_1 &= \frac{1}{f} \\
\Delta T_1 &= \frac{1}{f} \times (\frac{W-K_0+2P}{S_0}+1)
\end{split}
\end{equation}

\par Using the above formulas the delay line length for AlexNet is calculated in Table 1: \\


\begin{table}[h!]
{\centering \small
\begin{tabularx}{\textwidth}{
>{\hsize.6\hsize}
X
>{\hsize.5\hsize}
X
>{\hsize.5\hsize}
X
>{\hsize.2\hsize}
X
>{\hsize.2\hsize}
X
>{\hsize.3\hsize}
X
>{\hsize0.8\hsize}
X}

 \toprule

 & Dimension & Kernel Size & Stride & $\Delta t$ & $\Delta T$ & Delay Line Length \\ \midrule
1st-ConvLayer & $ 55 \times 55$ & $ 5 \times 5$ & $2$ & $\frac{1}{f}$ & $55 \times \frac{1}{f}$ & $4\times (55 + 1) \frac{1}{f} = \frac{224}{f}$ \\ 
2nd-ConvLayer & $ 27 \times 27$ & $3 \times 3$ & $2$ & $\frac{2}{f}$ & $55 \times \frac{2}{f}$ & $2\times (55 + 1) \frac{2}{f} = \frac{224}{f}$ \\ 
3rd-ConvLayer & $ 13 \times 13$ & $ 3 \times 3$ & $2$ & $\frac{4}{f}$ & $55 \times \frac{4}{f}$ & $2\times (55 + 1) \frac{4}{f} = \frac{448}{f}$ \\ 
4th-ConvLayer & $ 13 \times 13$ & $ 3 \times 3$ & $1$ & $\frac{8}{f}$ & $55 \times \frac{8}{f}$ & $2\times (55 + 1) \frac{8}{f} = \frac{896}{f}$ \\ 
5th-ConvLayer & $ 13 \times 13$ & $ 13 \times 13$ & NA & $\frac{8}{f}$ & $55 \times \frac{8}{f}$ & $12\times (55 + 1) \frac{8}{f} = \frac{5376}{f}$ \\ \bottomrule

\end{tabularx}}

\caption{AlexNet Delay Lines being fed to the system at frequency $f$.}

\end{table}

\newpage

\bibliography{sample}






\end{document}